\documentclass[prx,reprint,superscriptaddress,altaffillsymbol,amsmath]{revtex4-1}
\usepackage{amssymb}
\usepackage{epsfig}
 
\DeclareMathAlphabet{\pazocal}{OMS}{zplm}{m}{n}            
\usepackage{graphicx}
\usepackage{color}
\usepackage[dvipsnames]{xcolor}
\usepackage{bm}
\usepackage[normalem]{ulem}     
\usepackage{soul}
\usepackage{amsmath}
\usepackage{braket} 
\usepackage{rotating}
\usepackage{multirow} 
\usepackage{array}
 
\DeclareMathAlphabet{\pazocal}{OMS}{zplm}{m}{n}            

\newcommand{\co}{Cr$_2$O$_3$}
\newcommand{\Cro}{Cr$_2$O$_3$}
\newcommand{\alpara}{$\alpha_{\parallel}$}
\newcommand{\alperp}{$\alpha_{\perp}$}

\begin{document}
\title{On the sign of the linear magnetoelectric coefficient in \Cro{}}

\author{Eric Bousquet$^{*}$} 
\affiliation{University of Li\`{e}ge, Quartier Agora,
Allée du six Août 19, 4000 Li\`{e}ge 1, Belgium}
\thanks{Author names in alphabetical order, see author contribution statement}
\author{Eddy Leli\`{e}vre-Berna} 
\affiliation{Institut Laue Langevin, 71 Avenue des Martyrs, CS 20156, 38042 Grenoble, France}

\author{Navid Qureshi}
\affiliation{Institut Laue Langevin, 71 Avenue des Martyrs, CS 20156, 38042 Grenoble, France} 
\author{Jian-Rui Soh}
\affiliation{Laboratory for Quantum Magnetism, Institute of Physics, \'Ecole Polytechnique F\'ed\'erale de Lausanne, CH-1015 Lausanne, Switzerland} 
\author{Nicola A. Spaldin}
\affiliation{Materials Theory, ETH Zurich, Wolfgang-Pauli-Strasse 27, 8093 Zurich, Switzerland} 

\author{Andrea Urru}
\affiliation{Materials Theory, ETH Zurich, Wolfgang-Pauli-Strasse 27, 8093 Zurich, Switzerland} 
\affiliation{International School for Advanced Studies (SISSA), Via Bonomea 265, 34136 Trieste, Italy}
\author{Xanthe H. Verbeek}
\affiliation{Materials Theory, ETH Zurich, Wolfgang-Pauli-Strasse 27, 8093 Zurich, Switzerland} 
\email{xverbeek@ethz.ch}
\author{Sophie F. Weber}
\affiliation{Materials Theory, ETH Zurich, Wolfgang-Pauli-Strasse 27, 8093 Zurich, Switzerland} 
\date{\today}

\begin{abstract}
We establish the sign of the linear magnetoelectric coefficient, $\alpha$, in chromia, \Cro{}. \Cro{} is the prototypical linear magnetoelectric material, in which an electric (magnetic) field induces a linearly proportional magnetization (polarization), and a single magnetic domain can be selected by annealing in combined magnetic ($\mathbf{H}$) and electric ($\mathbf{E}$) fields. Opposite antiferromagnetic domains have opposite magnetoelectric responses, and which antiferromagnetic domain corresponds to which sign of response has previously been unclear. We use density functional theory (DFT) to calculate the magnetic response of a single antiferromagnetic domain of \Cro{} to an applied in-plane electric field at zero kelvin. We find that the domain with nearest neighbor magnetic moments oriented away from (towards) each other has a negative (positive) in-plane magnetoelectric coefficient, \alperp{}, at zero kelvin. We show that this sign is consistent with all other DFT calculations in the literature that specified the domain orientation, independent of the choice of DFT code or functional, the method used to apply the field, and whether the direct (magnetic field) or inverse (electric field) magnetoelectric response was calculated. Next, we reanalyze our previously published spherical neutron polarimetry data to determine the antiferromagnetic domain produced by annealing in combined $\mathbf{E}$ and $\mathbf{H}$ fields oriented along the crystallographic symmetry axis at room temperature. We find that the antiferromagnetic domain with nearest-neighbor magnetic moments oriented away from (towards) each other is produced by annealing in (anti-)parallel $\mathbf{E}$ and $\mathbf{H}$ fields, corresponding to a positive (negative) axial magnetoelectric coefficient, \alpara{}, at room temperature. Since \alperp{} at zero kelvin and \alpara{} at room temperature are known to be of opposite sign, our computational and experimental results are consistent.
\end{abstract}

\maketitle
\section{Introduction}
Materials in which both time-reversal $\Theta$ and space-inversion $\mathcal{I}$ symmetries are broken, while the product $\mathcal{I} \Theta $ symmetry is preserved, have a term in their free energy of the form
\begin{equation}
F(\mathbf{E},\mathbf{H})=-\frac{1}{V}\alpha_{ij}E_iH_j ,
 \label{eq:freeE_expand}
 \end{equation}
  where $\mathbf{E}$ / $\mathbf{H}$ are electric / magnetic fields, $\alpha$ is the nine-component {\it magnetoelectric} tensor (SI units $\mathrm{s/m}$) and $V$ is the unit cell volume. This term reveals two distinctive and related material properties. First, there is a preferred magnetic domain orientation, determined by the sign and form of $\alpha$, in simultaneous magnetic and electric fields, so that annealing in such a combination of fields, called {\it magnetoelectric annealing}, can be used to select for a specific magnetic domain. Second, by differentiating Eq. \ref{eq:freeE_expand} with respect to electric (magnetic) field to obtain the polarization (magnetization), we see that
 \begin{equation}
 P_i(\mathbf{E},\mathbf{H})=-\frac{\partial F}{\partial E_i}=\frac{1}{V}\alpha_{ij}H_j,
 \label{eq:ME_pol}
 \end{equation}
 and 
 \begin{equation}
M_i(\mathbf{E},\mathbf{H})=-\frac{1}{\mu_0}\frac{\partial F}{\partial H_i}=\frac{1}{\mu_0V}\alpha_{ji}E_i,
\label{eq:ME_M}
\end{equation}
where $\mu_0$ is the vacuum permeability. Eqs.~\ref{eq:ME_pol} and ~\ref{eq:ME_M} reveal a linear proportionality between an applied electric (magnetic) field and an induced magnetization $M_i$ (polarization $P_i$), with $\alpha$ the response tensor. 
Materials with non-zero $\alpha$ therefore show a linear magnetoelectric (ME) effect and are promising for spintronic applications since they enable voltage-control of magnetism~\cite{Borisov2008}. \\

\begin{figure}[b!]
    \centering
    \includegraphics{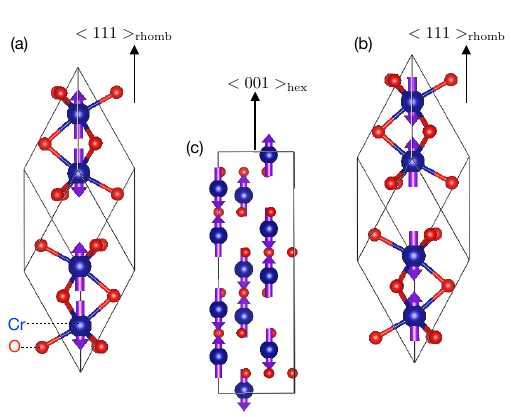}
    \caption{The crystal structure of \Cro{} showing the primitive rhombohedral unit cell, with the two AFM domains, the `out-pointing' domain (a) and the `in-pointing' domain (b). The goal of this work is to determine the absolute signs of the components of $\alpha$ for the two individual domains." The hexagonal setting, which we use in our experimental discussion, is shown in (c) for the in-pointing  magnetic domain. Note that $\langle001\rangle_{\mathrm{hex}}\parallel\langle111\rangle_{\mathrm{rhomb}}$.}
    \label{fig:Domains}
\end{figure} 
 
\indent Corundum-structure chromia, $\mathrm{Cr_2O_3}$, is the prototypical linear magnetoelectric, and the first material in which the linear ME effect was predicted~\cite{Dzyaloshinskii:1960} and measured~\cite{Astrov:1960,Astrov1961}. In addition to its historical relevance, $\mathrm{Cr_2O_3}$ has a high N\'{e}el temperature compared to other ME materials, and continues to be the primary material of focus in theoretical, experimental and technological studies of the ME effect. We show the primitive rhombohedral unit cell of $\mathrm{Cr_2O_3}$ in Fig.~\ref{fig:Domains}(a). Below its N\'{e}el temperature $T_\mathrm{N}$=\,307\,K~\cite{Brockhouse1953}, $\mathrm{Cr_2O_3}$ adopts a superexchange-mediated easy-axis antiferromagnetic (AFM) ``up-down-up-down" ordering of the magnetic dipole moments on the $d^3$ $\mathrm{Cr}^{3+}$ ions along the rhombohedral $\langle111\rangle$ direction~\cite{Dudko1971,Tobia2010}. The $\mathrm{R\bar{3}'c'}$ magnetic space group breaks both $\mathcal{I}$ and $\Theta$ while preserving $\mathcal{I} \Theta$, thus allowing a linear ME response~\cite{Fechner2018}. In Fig.~\ref{fig:Domains} (b) we show the primitive unit cell of the opposite AFM domain, with ``down-up-down-up'' magnetic dipole ordering. While (a) and (b) are energetically degenerate in the absence of external fields, they correspond to opposite ME domains. As a result, the signs of their linear ME responses are opposite, and they are obtained by ME annealing in opposite combinations of  $\mathbf{E}$ and $\mathbf{H}$ fields. In Fig.~\ref{fig:Domains}(c), we show the unit cell of $\mathrm{Cr_2O_3}$ in the hexagonal setting conventionally used in neutron diffraction, in which the hexagonal $\langle001\rangle$ axis is parallel to the rhombohedral $\langle111\rangle$ axis.  \\
 
\begin{figure}[b!]
    \centering
    \includegraphics{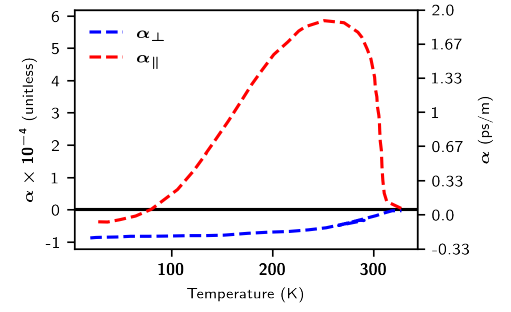}
    \caption{Temperature dependence of the parallel \alpara{} and perpendicular \alperp{} ME responses in \Cro{}, extracted from Ref. \cite{Astrov1961}. $\alpha$ is given in dimensionless units (by multiplying by the speed of light) multiplied by $10^{-4}$ (left-hand y axis) and SI $\mathrm{ps}/\mathrm{m}$ units (right-hand y axis). 
    }
    \label{fig:alphas_vs_T}
\end{figure}

\indent The symmetry of the $\mathrm{R\bar{3}'c'}$ magnetic space group allows for a diagonal response tensor $\alpha$, described by two independent components which we denote as $\alpha_{\parallel}$ and $\alpha_{\perp}$~\cite{Dzyaloshinskii:1960}:
 \begin{equation}
     \begin{pmatrix}
     \alpha_{\perp} & 0 & 0\\
     0 & \alpha_{\perp} & 0\\
     0 & 0 & \alpha_{\parallel}
     \end{pmatrix}.
     \label{eq:alpha_tensor}
 \end{equation}
 $\alpha_{\parallel}$ describes the magnetization (polarization) induced when $\mathbf{E}$ ($\mathbf{H}$) is applied along the rhombohedral $\langle111\rangle$ axis, and $\alpha_{\perp}$ refers to the perpendicular ME response when the field and induced property lie in the basal plane. Fig.~\ref{fig:alphas_vs_T} shows the measured temperature dependence of \alpara{} and \alperp{}, extracted from the original experimental report \cite{Astrov1961}. While \alperp{}, which results from the  $\mathbf{E}$-field induced canting of the magnetic dipole moments away from the easy axis~\cite{Iniguez:2008, blMalashevich2012}, follows the usual order-parameter onset below $T_\mathrm{N}$,  \alpara{} has a peak in magnitude just below $T_\mathrm{N}$ before decreasing and switching sign at low temperature. This is understood in terms of the response of spin fluctuations at high temperature~\cite{Mostovoy_et_al:2010}, with the orbital magnetization response~\cite{blMalashevich2012} dominating at low temperature. Importantly, at $T$=0\,K, relevant to first-principles calculations, \alperp{} and \alpara{} have the same sign, whereas at room temperature, relevant to many experimental setups, \alperp{} and \alpara{} have opposite signs.\\

\indent While the {\it relative} signs of \alperp{} and \alpara{} were established unambiguously in Ref.~\cite{Astrov:1960}, it was not possible at the time to determine which set of $\alpha$ values correspond to the out-pointing or in-pointing magnetic domains of Figs.~\ref{fig:Domains} a) and b). Instead, Ref.~\cite{Astrov:1960} showed that reversal of the AFM domain reverses the signs of $\alpha$ as required by symmetry, and that the measured magnitudes in multi-domain or poly-crystalline samples are substantially reduced due to cancellation effects. 
The experimental determination of the specific bulk AFM domain corresponding to a particular ME response is highly non-trivial and requires a generalized form of polarized neutron scattering called spherical neutron polarimetry; to our knowledge only four such experiment has been performed for Cr$_2$O$_3$~\cite{Brown/Forsyth/Tasset:1998,Brown/Forsyth/Tasset:1999,Brown_et_al:2002}. While in principle first-principles calculations based on density functional theory (DFT) yield this information directly, the AFM domain modeled is often not reported in the literature, and the sensitivity of the magnetic anisotropy to the details of the DFT parameters render an independent experimental determination desirable. To compound confusion, in both the theoretical and experimental literature the terms ``magnetic moments'' and ``spins'' have sometimes been used interchangeably, in spite of their being opposite in sign.\\

\indent The purpose of this paper is to establish unambiguously the signs of the ME effect corresponding to each of the two opposite AFM domains in Cr$_2$O$_3$. We achieve this goal by reviewing and reanalyzing the relevant computational and experimental literature, as well as presenting the results of our own new DFT calculations. In Sec.~\ref{sec:comp}, we begin by reviewing the DFT-based results for the zero-kelvin values of \alperp{} and \alpara{}, computed both by us and by others in earlier publications. We then perform a comprehensive cross-check of the domain-dependent sign of $\alpha$ using four different codes, three different methods for applying the external fields, and different choices of DFT parameters.

We find that the \emph{ab-initio} results give consistent signs for $\alpha$ across authors, DFT parameters, and codes used.\\
\indent In Sec.~\ref{sec:exp}, we reanalyze the seminal neutron polarimetry experiments which provided the first experimental indicator for the sign of $\alpha$~\cite{Forsyth_1988, Brown/Forsyth/Tasset:1998, Brown/Forsyth/Tasset:1999, Brown_et_al:2002}. While the stated conclusion of the original polarimetry papers contradicts the DFT findings, we show that this is actually due to the assumed sample orientation with respect to the instrument axes during analysis in Refs.~\cite{Forsyth_1988,Brown/Forsyth/Tasset:1999, Brown_et_al:2002}. When we account for and correct these inconsistencies, the raw polarimetry data indicate a room-temperature sign of $\alpha$ for a given AFM domain consistent with all DFT calculations (taking into account the experimental temperature dependence of $\alpha$ found by Astrov in Fig. \ref{fig:alphas_vs_T}). We hope that this paper clears up long-standing ambiguities and confusions in the literature, and facilitates future interpretations of theoretical and experimental data related to the ME effect in $\mathrm{Cr_2O_3}$ and other ME materials. 

\section{\label{sec:comp}Computational studies}

Several \textit{ab initio} studies of the magnitude and sign of the ME effect in \Cro{} have been performed previously \cite{blMalashevich2012, Iniguez:2008, Ye/Vanderbilt:2014, Bousquet/Spaldin/Delaney:2011, Mostovoy_et_al:2010}. Three main techniques have been employed: Explicit inclusion of i) a static magnetic field or ii) a static electric field within the DFT Hamiltonian, and iii) the so-called ``lattice-mediated'' method, in which a polar displacement of the ions simulates the application of an electric field. Both spin and orbital contributions to the response have been calculated, and $\alpha$ has been resolved into so-called \textit{clamped-ion} (the electronic response to an electric field with fixed ions) and \textit{lattice-mediated} (in which the ions are displaced by the electric field) components. 
Since most DFT codes (in particular \textsc{abinit}~\cite{Gonze:2020, Romero:2020}, \textsc{elk}~\cite{ELK}, \textsc{quantum espresso}~\cite{Gianozzi_et_al:2009, Giannozzi_et_al:2017} and \textsc{vasp}~\cite{Kresse/Furthmueller_CMS:1996,Kresse/Furthmueller_PRB:1996}) output magnetic moments rather than spins, we adopt this convention here. 

\begin{figure}[b!]
    \centering
    \includegraphics[width = 8cm]{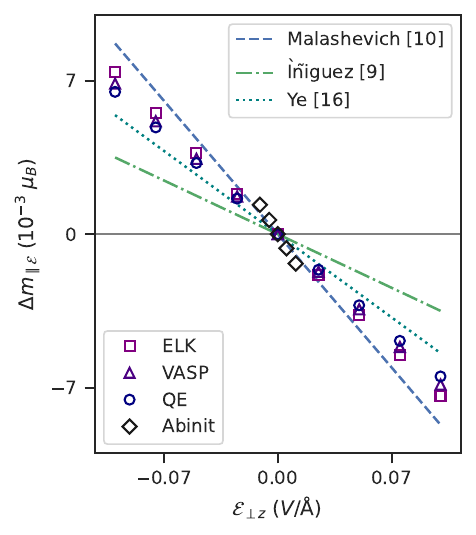}
    \caption{Induced net magnetic moment per rhombohedral unit cell parallel to an applied electric field oriented perpendicular to the easy axis,  as a function of electric field strength for the out-pointing domain. 
    The three lines show the response calculated from literature \alperp{} values. Markers indicate our results using four different DFT codes.}
    \label{fig:Mag_resp_diff_codes}
\end{figure}

\begin{table*}[t!]
\centering
\begin{tabular}{|c|c|c|c|c|c|c|c|c|c|c|c|c|}
 \cline{10-13}
    \multicolumn{1}{c}{} & \multicolumn{1}{c}{} & \multicolumn{1}{c}{} & \multicolumn{1}{c}{} & \multicolumn{1}{c}{} & \multicolumn{1}{c}{} & \multicolumn{1}{c}{} & \multicolumn{1}{c}{} & \multicolumn{1}{c}{} & \multicolumn{2}{|c|}{\alperp} & \multicolumn{2}{|c|}{\alpara} \\
\hline
 \multirow{2}{*}{Source} & \multirow{2}{*}{Code} & \multirow{2}{*}{PP} & \multirow{2}{*}{XC} & \multirow{2}{*}{U (eV)} & \multirow{2}{*}{SOC} & \multirow{2}{*}{Unit cell} & Contributions & \multirow{2}{*}{Method} & \multirow{2}{*}{ps/m} & g.u. &  \multirow{2}{*}{ps/m} & g.u. \\
 & & & & & & & to $\alpha$ & & & ($\times 10^{-4}$) & & ($\times 10^{-4}$) \\
 \hline
 Ref. \cite{blMalashevich2012} & QE & NC & PBE & no U & yes & PBE  & LM+CI, S+O & electric field & $-1.04$ & $-3.12$ &  $-0.02$ &  $-0.06$ \\
 \hline
 Ref. \cite{Iniguez:2008} & VASP & PAW & LDA & U$_{\text{eff}} = 2$ & yes & expt.  & LM, S & lattice-mediated & $-0.43$ & $-1.3$ &  $\boldsymbol{-}$ & $\boldsymbol{-}$  \\ 
 \hline
 Ref. \cite{Ye/Vanderbilt:2014} & QE & NC & PBE &  no U & yes &  PBE & LM, S+O & lattice-mediated & $-0.658$ & $-1.97$ & $-0.0221$  &  $-0.0663$\\
 \hline
 Ref. \cite{Bousquet/Spaldin/Delaney:2011} & VASP & PAW & LDA & U$_{\text{eff}} = 2$ & yes &  expt. & LM+CI,S  & Zeeman field & $-1.45$ & $-4.35$ & $\boldsymbol{-}$ & $\boldsymbol{-}$\\
 \hline
 \multirow{ 2}{*}{Ref. \cite{Mostovoy_et_al:2010}$^*$ } &  \multirow{ 2}{*}{VASP} &  \multirow{ 2}{*}{PAW} &  \multirow{ 2}{*}{LDA} &  \multirow{ 2}{*}{U$_{\text{eff}} = 2$} &  \multirow{ 2}{*}{no}  &  \multirow{ 2}{*}{expt.} &  \multirow{ 2}{*}{LM+CI, S}  &  magnetic  &  \multirow{ 2}{*}{$\boldsymbol{-}$} &  \multirow{ 2}{*}{$\boldsymbol{-}$} &  \multirow{ 2}{*}{$3.77$} &   \multirow{ 2}{*}{$11.3$} \\
 & & & & & & & & exchange & & & & \\
 \hline
 \multirow{ 2}{*}{This work} & \multirow{ 2}{*}{ELK} & \multirow{ 2}{*}{AE} & \multirow{ 2}{*}{LDA} & U = 4.0 & \multirow{ 2}{*}{yes} & \multirow{ 2}{*}{LDA} & \multirow{ 2}{*}{LM, S} & \multirow{ 2}{*}{lattice-mediated} & \multirow{2}{*}{$-0.921$} & \multirow{2}{*}{$-2.76$} &  \multirow{2}{*}{$\boldsymbol{-}$} & \multirow{2}{*}{$\boldsymbol{-}$} \\
 & & & & J = 0.5 & & & & & & & & \\
 \hline
 \multirow{ 2}{*}{This work} & \multirow{ 2}{*}{VASP} & \multirow{ 2}{*}{PAW} & \multirow{ 2}{*}{LDA} & U = 4.0 & \multirow{ 2}{*}{yes} & \multirow{ 2}{*}{LDA} & \multirow{ 2}{*}{LM, S} & \multirow{ 2}{*}{lattice-mediated} & \multirow{ 2}{*}{$-0.857$} & \multirow{ 2}{*}{$-2.57$} &  \multirow{ 2}{*}{$\boldsymbol{-}$} & \multirow{ 2}{*}{$\boldsymbol{-}$} \\
 & & & & J = 0.5 & & & & & & & & \\
 \hline
 This work & QE & US & PBE & no U &yes &  PBE & LM, S & lattice-mediated & $-0.773$ & $-2.32$ &  $\boldsymbol{-}$& $\boldsymbol{-}$\\
 \hline
 This work & Abinit & NC & LDA & no U  & yes & expt. & LM+CI, S & Zeeman field  &  $-1.48$ & $-4.44$ &  $\boldsymbol{-}$& $\boldsymbol{-}$ \\ 
 \hline
 This work & Abinit & NC & LDA & no U  & yes & expt. & LM+CI, S & electric field  &  $-1.48$ & $-4.44$ &  $\boldsymbol{-}$& $\boldsymbol{-}$ \\
 \hline
\end{tabular}
\caption{\label{tab:Diff_techniques} Overview of parameters used in different DFT calculations, performed with different codes (QE stands for \textsc{quantum espresso}). The short-hand notations for pseudopotentials (PP) and exchange-correlation functionals (XC) are: projected-augmented wave (PAW) \cite{Bloechl:1994}, norm-conserving  (NC), ultra-soft (US), all-electron (AE), local-density approximation (LDA) and generalized-gradient approximation with the Perdew-Burke-Ernzerhof (PBE) parametrization. SOC denotes spin-orbit coupling and the different contributions to $\alpha$ are indicated with LM  (lattice-mediated) and CI (clamped-ion), spin (S) and orbital (O).\\ $^*$ Results obtained at $T = 240\,$K.}
\end{table*}

First, we summarize the results of the various literature studies that report both the AFM domain studied and the sign of the calculated $\alpha$. The technical details for each calculation are summarized in Table~\ref{tab:Diff_techniques}. First, Malashevich \textit{et al.} \cite{blMalashevich2012} found \alpara{} and \alperp{} to have the same positive sign at 0 K for a domain with in-pointing moments. [as in Fig.~\ref{fig:Domains}(b)]. They used the finite electric-field method so that both spin and orbital contributions and the full lattice-mediated and electronic responses were included. For the same domain, \'I\~niguez~\cite{Iniguez:2008} used the 'lattice-mediated' method and obtained a positive 0\,K lattice-mediated spin ME response \alperp{}; since Ref.~\cite{Iniguez:2008} did not include orbital contributions, \alpara{} was zero. Also using the lattice-mediated approach but including the orbital contributions, Ye and Vanderbilt~\cite{Ye/Vanderbilt:2014} found positive \alpara{} and \alperp{} for the domain with in-pointing moments at 0\,K. Bousquet \textit{et al}.~\cite{Bousquet/Spaldin/Delaney:2011}, using an explicitly applied magnetic Zeeman field, including both the lattice-mediated and clamped-ion spin contributions, find a positive 0\,K \alperp{} for the in-pointing domain as well ~\cite{PriComm_delayney_2023}. Finally, Mostovoy \textit{et al.} \cite{Mostovoy_et_al:2010} considered the opposite domain (note that Fig. 1 of Ref.~\cite{Mostovoy_et_al:2010} shows \emph{spins}) and calculated the finite-temperature spin contribution to \alpara{}, using Monte-Carlo simulations of a DFT-derived model Hamiltonian containing Heisenberg exchanges and a magnetic moment - polarization coupling.
They found a positive \alpara{} in the temperature range of $T$ = 60-400\,K, consistent with a negative \alpara\,at $T$ = 0\,K [Fig.~\ref{fig:alphas_vs_T}]. Since their calculations modeled the out-pointing domain, these results are consistent with the other computational studies discussed earlier. 

To supplement the literature results, we perform a comprehensive cross-check of the domain-dependent sign of \alperp{} using four different codes and three different methods. First, we calculate the lattice-mediated spin contribution to \alperp{} using the lattice-mediated method, as described in Ref.~\cite{Iniguez:2008}, using the \textsc{elk} \cite{ELK}, \textsc{vasp}~\cite{Kresse/Furthmueller_CMS:1996, Kresse/Furthmueller_PRB:1996}, and \textsc{quantum espresso}~\cite{Gianozzi_et_al:2009, Giannozzi_et_al:2017} codes, with the parameters listed in Table~\ref{tab:Diff_techniques}. In all cases we find \alperp{}$<0$ for the out-pointing domain at 0\,K, consistent with the literature findings summarized above. In addition, we use the \textsc{abinit} code \cite{Gonze:2020, Romero:2020} to calculate the spin contribution to the ME effect by both explicitly applying an electric field as in Ref. \cite{blMalashevich2012}, and a magnetic Zeeman field method as in Ref. \cite{Bousquet/Spaldin/Delaney:2011}. Both methods give the same positive value of \alperp{} for the in-pointing domain at 0\,K. Computational details for the calculations in \textsc{elk}, \textsc{vasp}, \textsc{quantum espresso}, and \textsc{abinit} can be found in Appendices~\ref{sec:appELK}-\ref{sec:appAbinit}.
We list our calculated \alperp{} values in Table~\Ref{tab:Diff_techniques}, and in Fig. \ref{fig:Mag_resp_diff_codes}, we plot the induced in-plane magnetizations as a function of in-plane electric fields calculated here and from the literature. Although there is complete agreement on the sign of $\alpha$, it is clear that there is some spread in the magnitude of calculated values. This distribution cannot be explained only by the different contributions to $\alpha$ that were taken into account, and is  most likely also the result of the different choices in electronic structure code, electronic exchange parameters and convergence criteria. Considering these differences, the agreement on the magnitude of $\alpha$ is remarkable.

In summary, the calculated signs of $\alpha$ are consistent across DFT codes and methodologies, with the 0\,K  \alperp{} and \alpara{} positive for the in-pointing domain, the 0\,K \alperp{} negative for the out-pointing domain, and the room temperature \alpara{} positive for the out-pointing domain. We summarize this result in Tab. \ref{tab:DFT_result_summ} with the addition values at 0\,K and room T extrapolated. 

\begin{table}[t]
    \centering
    \begin{tabular}{m{0.3cm}|>{\centering}m{3cm}|>{\centering}m{2cm} | >{\centering\arraybackslash} m{2cm} |}
        \multirow{3}{*}{\rotatebox[origin=c]{90}{0 K}}
        &    &  \alpara &  \alperp \\
        & out-pointing domain & \textcolor{gray}{$-$}  & $\boldsymbol{-}$ \\
        & in-pointing domain & $\boldsymbol{+}$ & $\boldsymbol{+}$\\
        &         &                  &  \\

        \multirow{3}{*}{\rotatebox[origin=c]{90}{ RT }}
        &  & \alpara        & \alperp  \\  
        & out-pointing domain & $\boldsymbol{+}$ & \textcolor{gray}{$-$} \\
        & in-pointing domain & \textcolor{gray}{$-$} & \textcolor{gray}{$+$}       
    \end{tabular}
    \caption{Overview the sign of alpha for the two domains, at 0 K and room temperate (RT), as determined by different ab-initio calculations (black, bold), and extrapolated from the DFT results (grey) using the temperature dependence measured by Astrov, see Fig. \ref{fig:alphas_vs_T}.}
    \label{tab:DFT_result_summ}
\end{table}

\section{\label{sec:exp}Experimental studies}

To our knowledge, there exist four sets of data in which the magnetic structure of \co{} was measured using spherical neutron polarimetry (SNP), the generalized form of polarized neutron scattering~\cite{Forsyth_1988,Brown/Forsyth/Tasset:1998,Brown/Forsyth/Tasset:1999,Brown_et_al:2002}. This technique allows for both the detection of the domain imbalance between the two different magnetic structures shown in Fig.~\ref{fig:Domains}, and for the determination of the magnetic moment configuration of the predominant domain~\cite{Brown/Forsyth/Tasset:1998}. This is possible because with SNP, the polarization vectors of both the incident and scattered neutron beams are determined; in comparison, in conventional (uniaxial) polarized neutron scattering, the scattered neutron polarization information is only analyzed along the direction of the incident beam polarization~\cite{LelievreBerna:2007tm}. Therefore, SNP is an ideal method for elucidating which spin configuration shown in Fig.~\ref{fig:Domains} is stabilized by the parallel or anti-parallel combination of electric and magnetic fields.

The SNP measurements reported in Refs.~\cite{Forsyth_1988,Brown/Forsyth/Tasset:1998,Brown/Forsyth/Tasset:1999,Brown_et_al:2002} were performed at the IN20 and D3 beamlines at the Institut Laue Langevin (ILL, Grenoble), using the CRYOgenic Polarization Analysis Device (Cyopad). The Cryopad consists of a zero-magnetic field sample chamber surrounded by magnetic fields manipulating the incident (\textbf{P$_i$}) and scattered (\textbf{P$_f$}) beam polarizations~\cite{Tasset:1999wp,LelievreBerna:2005vj}. The field regions are decoupled with a pair of concentric superconducting Meissner shields combined with µ-metal yokes and screens. The incident neutron beam polarization was controlled using a combination of a nutator and precession coil, and was oriented along one of three orthogonal experimental co-ordinates which were defined as $x$, which is along the direction of the scattering vector $\textbf{Q}$, $z$, which is perpendicular to the horizontal scattering plane, and $y$, which completes the right-handed co-ordinate set. The polarization of the scattered neutron beam was also analyzed along these three principal axes using another set of precession and nutator coils.

In each of the four studies, the \co{} sample was aligned so that the crystal $b$-axis was perpendicular to the horizontal scattering plane, which allowed access to the $(h\,0\,l)$ reflections (importantly, this introduces an ambiguity between $b\|+z$ and $b\|-z$, which we will discuss in more detail in the following section). Here, the Miller indices correspond to the hexagonal setting of the rhombohedral ($R\bar{3}c$) unit cell of \co{} adopted in Refs.~\cite{Forsyth_1988,Brown/Forsyth/Tasset:1998,Brown/Forsyth/Tasset:1999,Brown_et_al:2002}. In the three most recent studies, prior to installing the sample in the Cryopad for the SNP measurements, the \co{} sample was cooled through the Néel temperature ($T_\mathrm{N}$$\sim$310\,K) in a combination of electric and magnetic fields oriented along the crystallographic $c$ axis to achieve an imbalance of 180$^\circ$ domain population~\cite{Brown/Forsyth/Tasset:1998,Brown/Forsyth/Tasset:1999,Brown_et_al:2002}. Brown \textit{et al.} reported that this annealing process stabilized a single AFM domain~\cite{Brown/Forsyth/Tasset:1998,Brown/Forsyth/Tasset:1999,Brown_et_al:2002}, and that the type of AFM domain (Fig.~\ref{fig:Domains}) could be chosen based on the relative orientation of the external magnetic (\textbf{H}) and electric fields (\textbf{E}). The experimental determination of which magnetic domain is favoured then boils down to the determination and interpretation of the sign of the polarization matrix element $P_{zx}$. 

Experimentally, $P_{zx}$ is determined by measuring two quantities, namely $n_{zx}$ and $n_{z\bar{x}}$, which are the number of scattered neutrons with the polarization parallel and antiparallel to $+x$ for the incident neutron polarization along $+z$. The experimental $P_{zx}$ matrix element is in turn obtained by taking the ratio,
\begin{align}
P_{zx} = \frac{n_{zx}-n_{z\bar{x}}}{n_{zx}+n_{z\bar{x}}},
\end{align}
for a given Bragg reflection \textbf{Q} = $(h\,k\,l)$. As such, the quantity $P_{zx}$ is bounded between -1 and 1. 

In order to determine which AFM domain is favoured, the authors in Refs.~\cite{Brown/Forsyth/Tasset:1999,Brown_et_al:2002,Brown/Forsyth/Tasset:1998,Forsyth_1988} expressed $P_{zx}$ in terms of three dimensionless quantities,
\begin{align}
P_{zx} = \eta \frac{-2 q_y \gamma}{1+\gamma^2}.
\end{align}

The $\eta$ term defines the population imbalance between the two magnetic domains, and is given by $\eta$ = ${(v^+-v^-)}/{(v^++v^-)}$, where $v^+$ and $v^-$ are the volumes of the two magnetic domains. Hence, the value of $\eta$ is bounded between 1 and -1. If the two magnetic domains are equally populated, the factor $\eta$ becomes 0. 
The term $q_y$ is determined by the orientation of the crystal with respect to the experimental set up, with the sign of $q_y$ depending on whether the crystallographic $+b$ axis is along the $+z$ or $-z$ direction of the experimental geometry; for example $q_y$ is +1 (-1) if the magnetic interaction vector $\textbf{M}_\perp(\textbf{Q})$ is parallel (anti-parallel) to the $+y$ 
axis of the experimental geometry. Hence, it is crucial to determine whether $+b$ is along the $+z$ or $-z$ direction. Finally, the term $\gamma$ is associated with the magnetic structure, with the sign of $\gamma$ being positive (negative) for the out-pointing (in-pointing) magnetic domain. 

Based on this discussion, we identify three inconsistencies across the four Refs.~\cite{Forsyth_1988,Brown/Forsyth/Tasset:1998,Brown/Forsyth/Tasset:1999,Brown_et_al:2002}, which we clarify here. (Note that the measurements in Refs.~\cite{Forsyth_1988,Brown/Forsyth/Tasset:1998,Brown/Forsyth/Tasset:1999,Brown_et_al:2002}
were made with the same crystal by the same group of coauthors so we expect the underlying physics to be consistent).

\subsection{Spin vs magnetic moment}

The first discrepancy is between Ref.~\cite{Brown_et_al:2002} and Ref.~\cite{Brown/Forsyth/Tasset:1998}, regarding the definition of spin and magnetic moments. In Ref.~\cite{Brown_et_al:2002}, the authors propose that the antiparallel \textbf{E} and \textbf{H} fields favor the `out-pointing' arrows [Fig.~\ref{fig:Domains}(a)] and designate the arrows as spin directions. On the other hand in Ref.~\cite{Brown/Forsyth/Tasset:1998}, the authors present `in-pointing' arrows [Fig.~\ref{fig:Domains}(b)], which they designate as magnetic moments, and state that this magnetic structure is stabilized by parallel \textbf{E} and \textbf{H} fields. Since the Cr spin direction and magnetic moment direction are anti-parallel, these two statements are incompatible.

In the neutron scattering community, however, the terms spin and magnetic moment are often used interchangeably to mean magnetic moment direction. We should therefore assume that the arrows in Ref.~\cite{Brown_et_al:2002} actually indicate magnetic moments, rather than spins as stated. This resolves the apparent discrepancy between Ref.~\cite{Brown/Forsyth/Tasset:1998} and Ref.~\cite{Brown_et_al:2002}.
\begin{figure}[t!]
 \centering
    \includegraphics[width = 9cm]{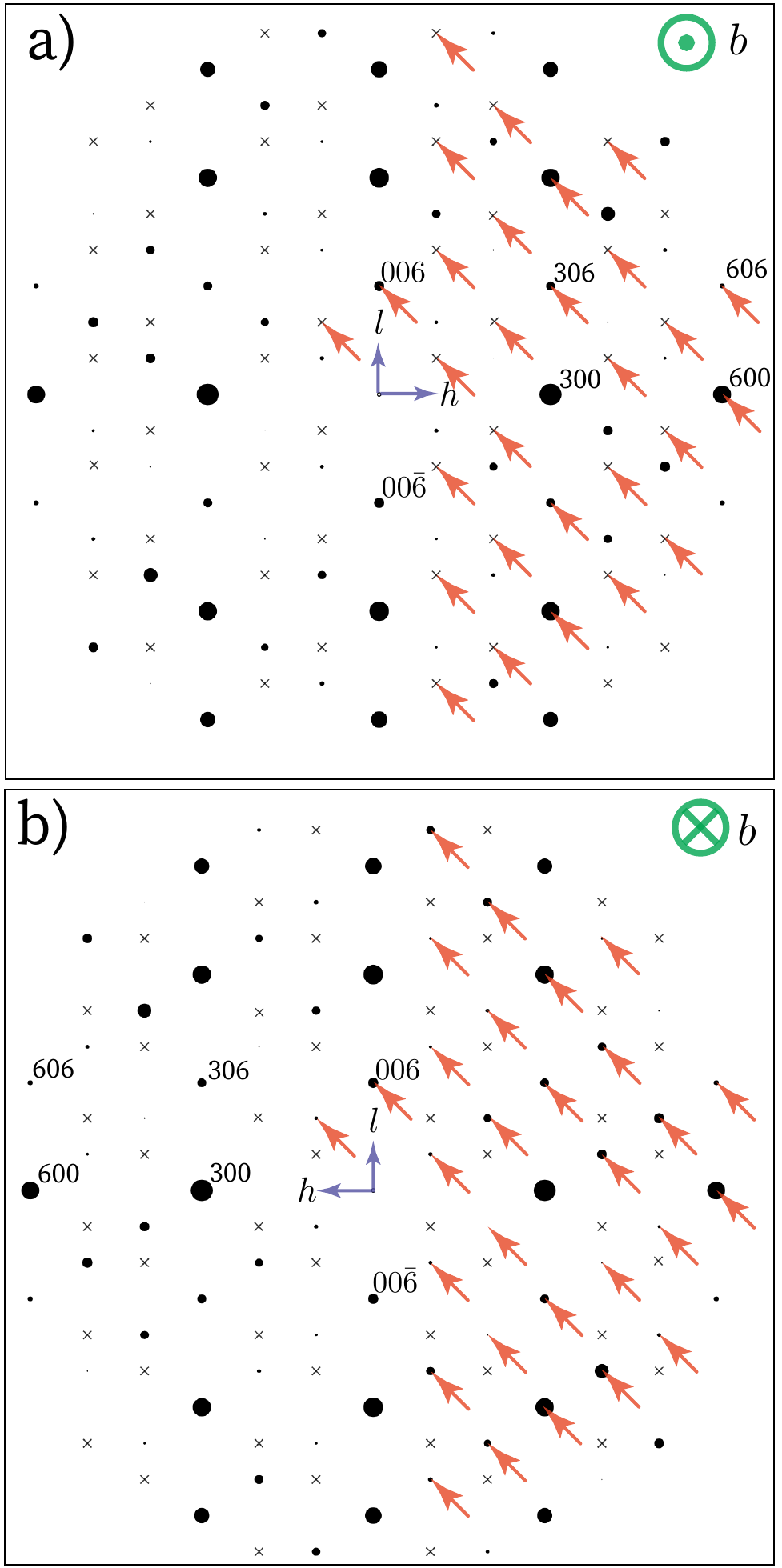}
     \caption{Plan view of the horizontal scattering plane of the reciprocal space map of \co{}, with the crystal $b$ axis (a) along the +$z$ direction quoted in~\cite{Forsyth_1988,Brown/Forsyth/Tasset:1999,Brown_et_al:2002} and (b) -$z$ directions, respectively. Here the filled circles indicate the allowed reflections, with the size of the circle proportional to the neutron scattering cross-section, whereas the cross ($\times$) denotes forbidden reflections. (a), The arrows indicate the reciprocal space location of the observed reflections in Refs.~\cite{Forsyth_1988,Brown/Forsyth/Tasset:1999,Brown_et_al:2002}, of which many are forbidden by the $R\bar{3}c$ space group of \co{}. (b), Instead, if the crystal $b$ axis were actually along the -$z$ direction, then the observed reflections denoted by the arrows can be accounted for.}
    \label{fig:Reciprocal}
\end{figure}
\subsection{Orientation of the crystal \textit{b} axis}

Second, the labelling of the Miller indices $(h0l)$ across the four reports is inconsistent. In the first report~\cite{Forsyth_1988}, the two reported reflections, namely $(1\,0\,2)$ and $(\bar{1}\,0\,4)$, are in fact forbidden by the $R\bar{3}c$ space group in the hexagonal setting of \co{}. In the subsequent study, the two reported reflections, $(\bar{1}\,0\,2)$ and $(1\,0\,\bar{2})$ are both allowed by $R\bar{3}c$. In the following two reports~\cite{Brown/Forsyth/Tasset:1999,Brown_et_al:2002}, where forty reflections were reported in total, thirty-two are in fact forbidden by the $R\bar{3}c$ space group of \co{}. The $h$ and $l$ Miller indices of the remaining eight reflections are both multiples of 3, e.g. $(3\,0\,6)$ and $(3\,0\,\bar{6})$, and are hence allowed.

\begingroup
\begin{table*}[t]
\begin{tabular}{|c | c |c|c|c|c c c|c c c|c c c|c c c|}
\multicolumn{17}{c}{Parallel \textbf{H} and \textbf{E}} \\
 \hline
         &             &                 &                    &                    & \multicolumn{3}{c|}{$P_{i}$}  &  \multicolumn{3}{c|}{$P_{f}^\mathrm{obs.}$} &  \multicolumn{3}{c|}{$P_{f}^\mathrm{calc.}$} &   \multicolumn{3}{c|}{$P_{f}^\mathrm{calc.}$}  \\[2pt]
Crystal  & $(h\,k\,l)$ &   Axis$\|z$     & Axis$\|\textbf{H}$ & Axis$\|\textbf{E}$ &    & &                        &        & &                                  &           \multicolumn{3}{c|}{out pointing}  &           \multicolumn{3}{c|}{in pointing}   \\
         &             &                 &                    &                    &  $x$&$y$&$z$                  &     $x$&$y$&$z$                             &     $x$  &$y$ & $z$                          &     $x$  &$y$ & $z$                                    \\
  \hline
 I & $\bar{1}\,0\,2$   & $0\,1\,0$       & $0\,0\,1$          & $0\,0\,1$          & \,\,\,\,0.00&0.00&0.88\,\,\,\,&\,\,\,\,0.83&0.06&0.08\,\,\,\,               & \,\,\,\,0.88&0.00&0.00\,\,\,\,               & \,\,\,\,-0.88&0.00&0.00\,\,\,\,\\ 
   &                   & $0\,\bar{1}\,0$ & $0\,0\,\bar{1}$    & $0\,0\,\bar{1}$    & \,\,\,\,0.00&0.00&0.72\,\,\,\,&\,\,\,\,-0.69&0.06&-0.06\,\,\,\,             & \,\,\,\,-0.72&0.00&0.00\,\,\,\,              & \,\,\,\,0.72&0.00&0.00\,\,\,\,\\
   &                   & $0\,\bar{1}\,0$ & $0\,0\,1$          & $0\,0\,1$          & \,\,\,\,0.00&0.00&0.72\,\,\,\,&\,\,\,\,-0.70&0.05&-0.05\,\,\,\,             & \,\,\,\,-0.72&0.00&0.00\,\,\,\,               & \,\,\,\,0.72&0.00&0.00\,\,\,\,\\[5pt]  
   \hline
 II&  $1\,0\,\bar{2}$ & $0\,\bar{1}\,0$ & $0\,0\,\bar{1}$    & $0\,0\,\bar{1}$    & \,\,\,\,0.88&0.00&0.00\,\,\,\,&\,\,\,\,-0.10&0.00&0.86\,\,\,\,              & \,\,\,\,0.00&0.00&0.88\,\,\,\,               & \,\,\,\,0.00&0.00&-0.88\,\,\,\,\\ 
   &                   & $0\,\bar{1}\,0$ & $0\,0\,\bar{1}$    & $0\,0\,\bar{1}$    & \,\,\,\,0.00&0.88&0.00\,\,\,\,&\,\,\,\,0.06&0.88&0.03\,\,\,\,               & \,\,\,\,0.00&0.88&0.00\,\,\,\,               & \,\,\,\,0.00&0.88&0.00\,\,\,\,\\ 
   &                  & $0\,\bar{1}\,0$ & $0\,0\,\bar{1}$    & $0\,0\,\bar{1}$    & \,\,\,\,0.00&0.00&0.88\,\,\,\,&\,\,\,\,-0.87&0.03&0.02\,\,\,\,              & \,\,\,\,-0.88&0.00&0.00\,\,\,\,               & \,\,\,\,0.88&0.00&0.00\,\,\,\,\\ [5pt]
   
   \hline
\end{tabular}
\caption{\label{tab:Pol_Mat_para} Comparison between the measured ($P_{f}^\mathrm{obs.}$) and calculated ($P_{f}^\mathrm{calc.}$) polarization matrices with `out-pointing' and `in-pointing' magnetic domain for the data collected with \textbf{E} and \textbf{H} parallel. The data is consistent with the `out-pointing' magnetic domain, as shown in Fig.~\ref{fig:Domains}(a). }
\end{table*}
\begin{table*}
\begin{tabular}{|c | c |c|c|c|c c c|c c c|c c c|c c c|}
\multicolumn{17}{c}{Anti-parallel \textbf{H} and \textbf{E}} \\
 \hline
         &             &                 &                    &                    & \multicolumn{3}{c|}{$P_{i}$}  &  \multicolumn{3}{c|}{$P_{f}^\mathrm{obs.}$} &  \multicolumn{3}{c|}{$P_{f}^\mathrm{calc.}$} &   \multicolumn{3}{c|}{$P_{f}^\mathrm{calc.}$}  \\[2pt]
Crystal  & $(h\,k\,l)$ &   Axis$\|z$     & Axis$\|\textbf{H}$ & Axis$\|\textbf{E}$ &    & &                        &        & &                                  &           \multicolumn{3}{c|}{out pointing}  &           \multicolumn{3}{c|}{in pointing}   \\
         &             &                 &                    &                    &  $x$&$y$&$z$                  &     $x$&$y$&$z$                             &     $x$  &$y$ & $z$                          &     $x$  &$y$ & $z$                                    \\
  \hline
I  &  $\bar{1}\,0\,2$  & $0\,\bar{1}\,0$ & $0\,0\,1$          & $0\,0\,\bar{1}$    & \,\,\,\,0.00&0.00&0.72\,\,\,\,&\,\,\,\,0.71&0.12&0.02\,\,\,\,               & \,\,\,\,-0.72&0.00&0.00\,\,\,\,               & \,\,\,\,0.72&0.00&0.00\,\,\,\,\\ 
   &                   & $0\,\bar{1}\,0$ & $0\,0\,\bar{1}$    & $0\,0\,1$          & \,\,\,\,0.00&0.00&0.72\,\,\,\,&\,\,\,\,0.70&0.16&0.05\,\,\,\,               & \,\,\,\,-0.72&0.00&0.00\,\,\,\,               & \,\,\,\,0.72&0.00&0.00\,\,\,\,\\ [5pt]
   \hline
II & $1\,0\,\bar{2}$   & $0\,\bar{1}\,0$ & $0\,0\,\bar{1}$    & $0\,0\,1$          & \,\,\,\,0.88&0.00&0.00\,\,\,\,&\,\,\,\,0.12&0.00&-0.85\,\,\,\,              & \,\,\,\,0.00&0.00&0.88\,\,\,\,               & \,\,\,\,0.00&0.00&-0.88\,\,\,\,\\ 
      &                   & $0\,\bar{1}\,0$ & $0\,0\,\bar{1}$    & $0\,0\,1$          & \,\,\,\,0.00&0.88&0.00\,\,\,\,&\,\,\,\,0.00&0.88&-0.05\,\,\,\,              & \,\,\,\,0.00&0.88&0.00\,\,\,\,               & \,\,\,\,0.00&0.88&0.00\,\,\,\,\\
      &                   & $0\,\bar{1}\,0$ & $0\,0\,\bar{1}$    & $0\,0\,1$          & \,\,\,\,0.00&0.00&0.88\,\,\,\,&\,\,\,\,0.86&0.12&0.14\,\,\,\,               & \,\,\,\,-0.88&0.00&0.00\,\,\,\,               & \,\,\,\,0.88&0.00&0.00\,\,\,\,\\[5pt]
    \hline
II & $\bar{1}\,0\,2$   & $0\,\bar{1}\,0$ & $0\,0\,\bar{1}$    & $0\,0\,1$          & \,\,\,\,0.88&0.00&0.00\,\,\,\,&\,\,\,\,0.10&0.06&-0.86\,\,\,\,              & \,\,\,\,0.00&0.00&0.88\,\,\,\,               & \,\,\,\,0.00&0.00&-0.88\,\,\,\,\\
   &                   & $0\,\bar{1}\,0$ & $0\,0\,\bar{1}$    & $0\,0\,1$          & \,\,\,\,0.00&0.88&0.00\,\,\,\,&\,\,\,\,-0.09&0.88&0.02\,\,\,\,              & \,\,\,\,0.00&0.88&0.00\,\,\,\,               & \,\,\,\,0.00&0.88&0.00\,\,\,\,\\ 
   &                   & $0\,\bar{1}\,0$ & $0\,0\,\bar{1}$    & $0\,0\,1$          & \,\,\,\,0.00&0.00&0.88\,\,\,\,&\,\,\,\,0.87&0.03&0.08\,\,\,\,               & \,\,\,\,-0.88&0.00&0.00\,\,\,\,               & \,\,\,\,0.88&0.00&0.00\,\,\,\,\\[5pt]    
   
    \hline
\end{tabular}
\caption{\label{tab:Pol_Mat_anti_para} Comparison between the measured ($P_{f}^\mathrm{obs.}$) and calculated ($P_{f}^\mathrm{calc.}$) polarization matrices with `out-pointing' and `in-pointing' magnetic domain for the data collected with \textbf{E} and \textbf{H} anti-parallel. The data is consistent with the `in-pointing' magnetic domain, as shown in Fig.~\ref{fig:Domains}(b). }
\end{table*}
\endgroup

Given that the magnetic propagation vector of \co{} is $\textbf{Q}_\mathrm{m}$=$(0\,0\,0)$, the magnetic scattering intensity occurs at the same reciprocal space location as the structural Bragg peaks of \co{}. As such, the Miller index of the magnetic/nuclear reflections should follow the general condition of the $R\bar{3}c$ space group, where $-h+k+l$=$3n$. Since the four reports were concerned with reflections in the $(h\,0\,l)$ plane, the observed reflections should obey the rule $-h+l$=$3n$, given that $k$=0. In figure~\ref{fig:Reciprocal}(a) we plot the calculated reciprocal space maps for \co{} in the $(h\,0\,l)$ scattering plane, assuming that the $+b$ crystal axis is along the $+z$ direction as stated in the original papers. Here, the allowed reflections, such as $(\bar{1}\,0\,2)$ and $(1\,0\,4)$, are denoted by the black filled circles, and the reciprocal space location of the forbidden reflections that do not obey $-h+l$=$3n$ are shown by the crosses ($\times$). 

The observed reflections in Refs.~\cite{Forsyth_1988,Brown/Forsyth/Tasset:1999,Brown_et_al:2002} are denoted by the arrows in the reciprocal space map. Indeed, many of the observed reflections, including $(1\,0\,2)$ and $(1\,0\,\bar{4})$, are in fact forbidden by the $R\bar{3}c$ space group.

If instead, we assume that the $+b$ crystal axis was oriented along the $-z$ direction (rather than $+z$), then the reciprocal space location of all forty-two observed reflections reported in~\cite{Forsyth_1988,Brown/Forsyth/Tasset:1999,Brown_et_al:2002}, is fully compatible with the $R\bar{3}c$ space group. This scenario is very plausible, due to a possible mix up between the $+b$ and $-b$ crystal axes, which are in-equivalent in \co{}. As shown in Fig.~\ref{fig:Reciprocal}(b), where we plot the calculated reciprocal space maps for \co{} in the $(h\,0\,l)$ scattering plane, assuming that the $+b$ crystal axis is along the $-z$ direction, the reciprocal space location of the observed reflections denoted by all of the arrows can now be accounted for.

Changing the direction of the $+b$ axis has two main consequences for the interpretation of the results in Refs~\cite{Forsyth_1988,Brown/Forsyth/Tasset:1999,Brown_et_al:2002}. First it swaps the $h$ Miller index of the reflections, such that the observed peaks which were designated as $(h\,0\,l)$ should be assigned as $(\bar{h}\,0\,l)$ instead. This would allow the thirty-four reflection which are originally forbidden now be compatible with $R\bar{3}c$, i.e. to obey the $-h+l$=$3n$ condition. The remaining eight reflections which have the $h$ and $l$ Miller indices both being multiples of 3, still obey this condition. The second is that in Ref~\cite{Brown/Forsyth/Tasset:1998} the sign of $q_y$ changes, which means that the interpretation of which magnetic domain is favoured also changes. 

Therefore, we conclude that the conjugate field with \textbf{E} and \textbf{H} parallel favours the `out-pointing' domain, as shown in Fig.~\ref{fig:Domains}(a). By the same token, the antiparallel \textbf{E} and \textbf{H} field favours the `in-pointing' domain. This is opposite to the interpretation in Ref.~\cite{Brown_et_al:2002}.

\subsection{Sign of $\gamma$}

Finally, the third inconsistency is between Refs.~\cite{Brown/Forsyth/Tasset:1999} and~\cite{Brown_et_al:2002}. In these studies, the $\gamma$ term was obtained by measuring the polarization $P_{zx}$ component of various reflections. Ref.~\cite{Brown/Forsyth/Tasset:1999} reports, in Table 3, the $\gamma$ values for twelve reflections obtained on the IN20 instrument with thermal neutrons ($\lambda$=1.532 \AA{}). On the other hand, Ref.~\cite{Brown_et_al:2002} reports the measurements of $\gamma$ for a further fifteen reflections acquired on the D3 instrument with hot neutrons. Table 2 of Ref.~\cite{Brown_et_al:2002} lists the $\gamma$ data acquired from the D3 instrument along with those measured on the IN20 instrument, which were reported in~\cite{Brown/Forsyth/Tasset:1999}.

The discrepancy lies in the sign of $\gamma$ of the data collected on the IN20 instrument, which are reported both in Table 2 of Ref.~\cite{Brown_et_al:2002} and also in Table 3 of Ref.~\cite{Brown/Forsyth/Tasset:1999}. Although the Miller indices of the twelve reflections and their corresponding magnitude of $\gamma$ are the same, the signs are different. Since the sign of $\gamma$ is used to interpret whether the magnetic domain is `out-pointing' or `in-pointing', this discrepancy calls into question which sign of $\gamma$ was measured.

To resolve the ambiguity, here we use the  \textsc{Mag2Pol} software~\cite{Mag2pol} to re-analyze the measured spherical neutron polarimetry data presented in Table 2 of Ref.~\cite{Brown/Forsyth/Tasset:1998}. We choose this data set because the Miller indices are allowed by the $R\bar{3}c$ space group, and the raw data are presented explicitly. Moreover, these measurements were performed on cooling the \Cro{} sample with a conjugate field of parallel or anti-parallel \textbf{E} and \textbf{H} fields through $T_\mathrm{N}$ to $T$=290\,K, where the measurements were performed. 
Tables~\ref{tab:Pol_Mat_para} and~\ref{tab:Pol_Mat_anti_para} tabulate the measured polarization matrices for the case where \textbf{E} and \textbf{H} are parallel and anti-parallel, respectively, along with the results of our new analysis 
for the two cases where the magnetic domain is `out-pointing' or `in-pointing'.

Our analysis assuming an `out-pointing' domain is consistent with the measured scattered neutron polarization for the case where \textbf{E} and \textbf{H} are parallel, contrary to the conclusions in Refs.~\cite{Forsyth_1988,Brown_et_al:2002,Brown/Forsyth/Tasset:1998,Brown/Forsyth/Tasset:1999}. Similarly, for the case where \textbf{E} and \textbf{H} are anti-parallel, we find that the measured polarization matrices are consistent with an `in-pointing' domain. 

\section{Conclusion}
\begin{figure*}[t]
  \centering
   \includegraphics[width= 15 cm]{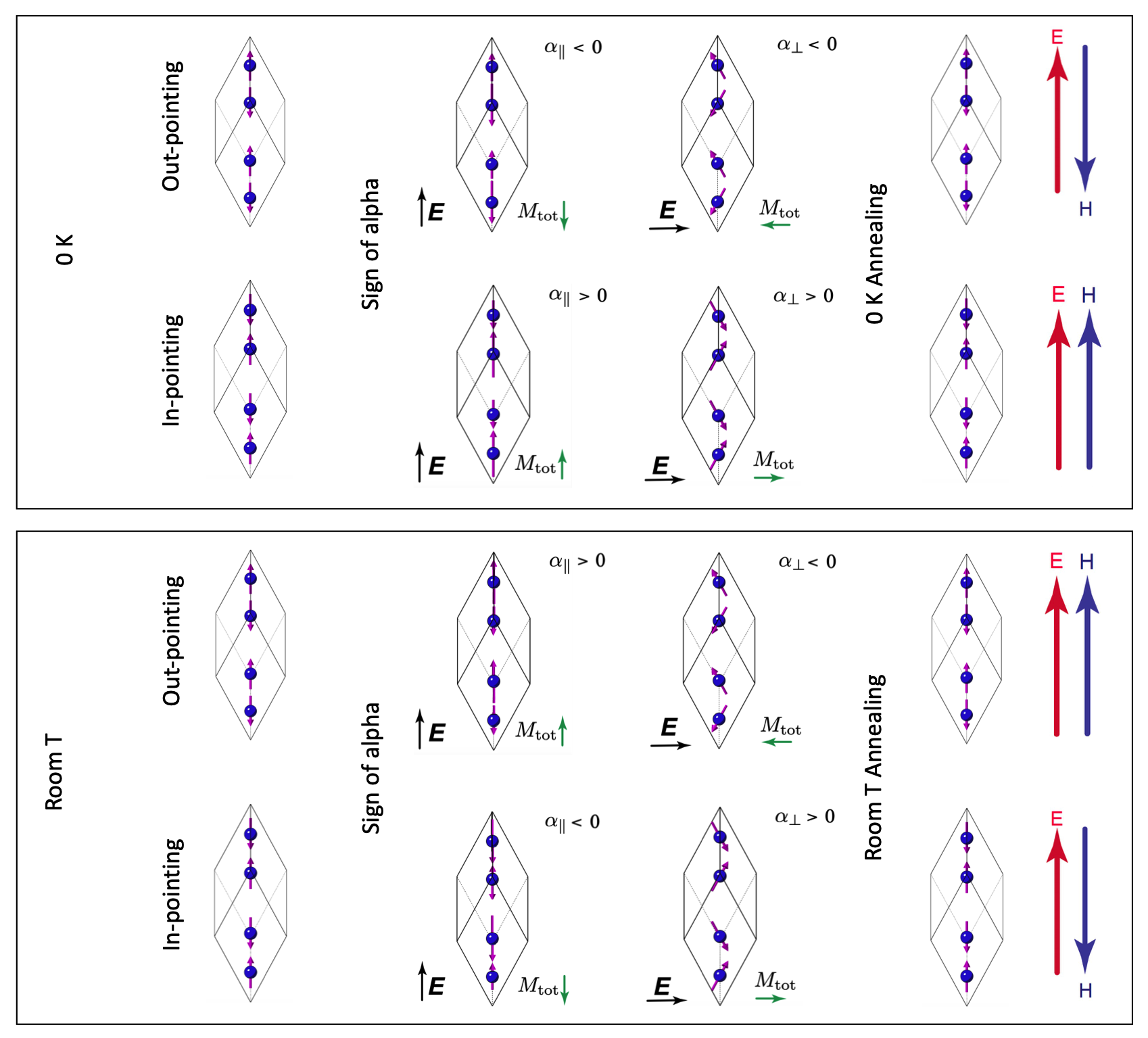}
 \caption{The in- and out-pointing domains of \Cro{} with the sign of \alperp{} and \alpara{} and alignment of the $\mathbf{E}$ and $\mathbf{H}$ fields during the ME annealing that favours each domain at 0 K and at room temperature. }
 \label{fig:summary_fig} 
\end{figure*}

\indent We have combined a literature review, new \emph{ab-initio} results, and a careful reanalysis of spherical neutron polarimetry data in an effort to resolve long-standing confusion regarding the domain-dependent sign of the ME coefficient in $\mathrm{Cr_2O_3}$. We have shown that all \emph{ab-initio} results to date are in agreement in the assignments of negative and positive low-temperature $\alpha$ to the out-pointing and in-pointing domains depicted in Fig.~\ref{fig:Domains}. These conclusions are remarkably consistent across multiple codes and methods. Gratifyingly, the room-temperature spherical neutron polarimetry \emph{data} are consistent with the low-temperature \emph{ab-initio} findings given that the room-temperature sign of \alpara{} is opposite to its low-temperature sign. The opposite interpretation in some of our literature experimental papers stems from a sign error due to subtle inconsistencies in the analysis which we discussed in Sec. \ref{sec:exp}. The confusion and deceptive inconsistency have also been compounded in the past by ambiguous terminology from numerous authors related to the usage of ``spin" versus ``magnetic" moment. We summarize the relationship between the domains and the sign of $\alpha$, as well as the necessary alignment of the $\mathbf{E}$ and $\mathbf{H}$ fields during magnetoelectric annealing, in Fig. \ref{fig:summary_fig}.

We mention here an important consequence of our work for a related feature of $\mathrm{Cr_2O_3}$; the magnitude and sign of the uncompensated magnetization on the $(001)$ surface for a given bulk domain \cite{He_et_al:2010,Borisov2016}. At the $0$-$\mathrm{K}$ limit in the absence of thermal fluctuations, the direction of the $(001)$ surface magnetization is unambiguously determined by the bulk domain which is selected in the ME annealing process. For example, with the in-pointing domain depicted for the hexagonal cell in Fig. \ref{fig:Domains}(c), the surface magnetization from the dangling $\mathrm{Cr}$ at the $(001)$ surface points outwards (positive). However, for any experimental characterizations performed at room-temperature, the relation between the bulk domain and the sign of surface magnetization is much less clear. Indeed, recent DFT-Monte Carlo calculations performed by some of the authors \cite{Weber2023} indicate that the $(001)$ surface magnetic moments of $\mathrm{Cr_2O_3}$ are essentially paramagnetic at room temperature due to weak coupling to the bulk order parameter. Thus, it is likely that for a fixed domain, the surface magnetization is substantially reduced, or even switches sign, with respect to its $0$ $\mathrm{K}$ value. Now that we have definitively determined which domain is selected by a given ME annealing at room temperature, it will be very interesting to re-examine, and perform new, experimental measurements of surface magnetization to determine its sign for an unambiguous selection of bulk domain.

We hope that our work convincingly demonstrates the previously questioned consistency of computational and experimental findings on the sign of the ME coefficient in $\mathrm{Cr_2O_3}$, and that it may motivate new, updated polarimetry measurements to test and confirm existing experimental and theoretical results. We also hope that this paper will assist in the correct interpretation of future studies of $\mathrm{Cr_2O_3}$, as well as providing a cautionary tale for similar investigations of other ME materials.

\section{Acknowledgements}
JRS, NAS, AU, XHV, and SFW were supported by the ERC under the European Union’s Horizon 2020 research and innovation programme grant No. 810451 and by the ETH Z\"urich. Computational resources for the \textsc{elk} and \textsc{vasp} calculations were provided by ETH Z\"urich’s Euler cluster, and by SISSA through its Linux Cluster and ITCS for the \texttt{Quantum ESPRESSO} calculations. EB acknowledges the FNRS for support and the computational resources provided by the Consortium des \'Equipements de Calcul Intensif (C\'ECI, FNRS grant No. 2.5020.11) and the Tier-1 supercomputer of the F\'ed\'eration Wallonie-Bruxelles funded by the Walloon Region (Grant No. 1117545). The authors thank Kris Delaney for providing us with the original input files for Ref.~\cite{Bousquet/Spaldin/Delaney:2011} so that we could extract the magnetic domain used for the applied Zeeman field calculations.
ELB and NQ would like to pay tribute to their friend F. Tasset, inventor of the Cryopad, who passed away earlier this year.

\section{Data availability statement}
All data that support the findings of this study are included
within the article (and any supplementary files).

\section{Author Contributions}
XHV performed the calculations in \textsc{elk} and \textsc{vasp}, AU performed those in \texttt{Quantum ESPRESSO} and EB those in \textsc{abinit}. JRS, NQ and ELB performed the re-analysis of the SNP data. NAS conceived of and coordinated the project.
All authors co-wrote the manuscript.

\section{Conflict of interest}
The authors declare that there is no conflict of interest.

\appendix{}

\section{\label{sec:appELK}Computational details ELK}
Our DFT calculations in the augmented-plane wave (APW) code \textsc{elk} were performed with spin-orbit interaction included, using the non-collinear local spin density approximation (LSDA) \cite{Perdew/Zunger:1981}. Correlation effects were taken into account by applying a rotationally invariant Hubbard U correction \cite{Liechtenstein_U} on the Cr $d$ states, with U $=4.0$ eV and J $=0.5$ eV, which well describe the physics of \Cro{} \cite{shi:2009, mu:2014, Fechner2018}. Muffin-tin spheres were used to describe the Cr and O core states, with radii of 1.0716 \AA \space and 0.80435 \AA. These radii are reduced by 4\% with respect to the standard setting to prevent overlap of the muffin-tin spheres. The APW functions and the potential were expanded in a spherical harmonics basis, with cut-offs $l_{\text{max} (\text{apw})} = l_{\text{max} (\text{o})} = 12$. A $6 \times 6 \times 6$ $\Gamma$-centered k-point mesh was used to sample the Brillouin Zone (BZ) \cite{Monkhorst/Pack:1976}. 
We obtained the spin contributions to the lattice-mediated ME response in the xy-plane using the lattice-mediated method of Ref. \cite{Iniguez:2008}, in which the response is constructed from a superposition of the magnetic moments induced by freezing in those eigenmodes of the force constant matrix that give a net polarization, in this case those with $E_u$ symmetry. We used LSDA + U relaxed lattice parameters and atomic positions obtained from \textsc{vasp} calculations (see the description below). Force constant matrix eigenmodes and their energies were obtained from \textsc{vasp} interfaced with phonopy \cite{phonopy, phonopy-phono3py-JPSJ}. Born effective charges, used to calculate the polarization, were taken from \textsc{vasp} calculations as well. 

\section{\label{sec:appVASP}Computational details VASP}
In the plane-wave code \textsc{vasp}, we performed density functional theory calculations with the LSDA+U method, spin-orbit coupling included, and a Hubbard U correction on the Cr $d$ states, with U (J) = $4.0$ ($0.5$) eV, as in the \textsc{elk} calculations. The ionic cores of Cr and O were described with projector-augmented wave pseudopotentials \cite{Bloechl:1994}. We used the following settings for the valence electrons: Cr 3p$^6$3d$^5$4s$^1$ and O 2s$^2$2p$^4$, corresponding to the datasets Cr\_sv and O. We used a kinetic energy cut-off of 800 eV for the wavefunctions and performed the BZ integrations using a uniform $\Gamma$-centered $7\times7\times7$ k-point mesh. Structural and electronic relaxations performed with these parameters yielded a band gap and magnetic moment close to known experimental values and lattice parameters of $a = 5.31 \, \mathrm{\AA}$, the length of the rhombohedral unit cell vectors and $\alpha = 54.87^{\circ} $, the angle between the unit cell vectors. These values are 0.78\% and 0.26\% smaller than experiment \cite{Hill:2010}. As for the \textsc{elk} calculations, we used the method of Ref.~\cite{Iniguez:2008} to construct the lattice-mediated spin response to an applied electric field, from the net spin magnetic moment induced by freezing in appropriate eigenmodes of the force constant matrix. The eigenmodes and corresponding energies were calculated by interfacing \textsc{vasp} with phonopy. The polarizations of each of the eigenmodes were obtained from the product of the atomic displacements of the mode and the Born effective charges $Z^{e}$. We computed the $Z^{e}$ by displacing each atom in the unit cell along each Cartesian direction and determining the ionic polarization using the modern theory of polarization, as implemented in \textsc{vasp} in the LCALCPOL routine. These calculations were performed for four displacements of different magnitudes, allowing us to assess the linear response regime. The final $Z^{e}$ were obtained from the average of the $Z^{e}$ for different atoms of the same species and different displacements within the linear regime. 
 
\section{\label{sec:appQE}Computational details Quantum Espresso}

First-principles calculations in \texttt{Quantum ESPRESSO} \cite{Gianozzi_et_al:2009, Giannozzi_et_al:2017} and \texttt{thermo\_pw} \cite{thermo_pw} were performed in non-collinear DFT using the generalized gradient approximation, with the Perdew-Burke-Ernzerhof parametrization of the exchange-correlation energy \cite{Perdew/Burke/Ernzerhof:1996}. Ions were described by fully relativistic ultrasoft pseudopotentials (PPs) \cite{US_Vanderbilt}, with 3$s$, 3$p$, 4$s$, and 4$d$ valence electrons for Cr (PP \texttt{Cr.rel-pbe-spn-rrkjus\_psl.0.2.3.UPF} from pslibrary 1.0.0 \cite{pslibrary,pslibrary_2}) and with 2$s$ and 2$p$ valence electrons for O (PP \texttt{O.rel-pbe-n-rrkjus\_psl.0.1.UPF} from pslibrary 0.1). The pseudo wavefunctions (charge density) were expanded in a plane-wave basis set with kinetic energy cut-off of 140 (560) Ry. BZ integrations were performed using a shifted $\textbf{k}$-points mesh of $6 \times 6 \times 6$ points. The lattice-mediated spin contribution to the ME response was computed following the approach of Ref. \cite{Iniguez:2008}: specifically, Born effective charges and phonon frequencies at $\Gamma$ were computed using density functional perturbation theory \cite{DFPT_review_Baroni}. 

\section{\label{sec:appAbinit}Computational details Abinit}
The ABINIT calculations (version 8.8) were done with the norm-conserving pseudo-potentials coming from the PseudoDojo project~\cite{pseudodojo} (v0.3) and within the LDA approximation for the exchange correlation functional without Hubbard U correction.
We used a kinetic energy cut-off of 40 Ha (1088 eV) for the plane-wave expansion and integrated the BZ using a Monkhorst-Pack $\mathbf{k}$-points mesh of $3\times 3\times 3$ points, shifted by (0.5, 0.5, 0,5).
Spin-orbit coupling was included in all the calculations for both applied Zeeman field and applied electric field calculations.
The cell parameters and shape were fixed to the experimental ones ($a = $ 5.37 \AA\ and $\alpha = 55.13^{\circ}$).
The forces were relaxed up to a tolerance of $2.7 \times 10^{-8}$ eV/\AA\ and the SCF cycles to a tolerance of $2.7 \times 10^{-9}$ eV/\AA\ on the force residual.
 
\bibliographystyle{unsrt}
\bibliography{References.bib,sign_of_alpha.bib,SNP.bib,references_Andrea.bib}

\begin{thebibliography}{10}

\bibitem{Borisov2008}
Borisov P, Hochstrat A, Shvartsman~V V, Kleemann W, and Hauck~P M.
\newblock Magnetoelectric {C}r$_2${O}$_3$ for spintronic applications.
\newblock {\em Integr. Ferroelectr.}, 99:69--76, 2008.

\bibitem{Dzyaloshinskii:1960}
Dzyaloshinskii~I E.
\newblock On the magneto-electrical effect in antiferromagnets.
\newblock {\em Sov. Phys. JETP}, 10:628--629, 1960.

\bibitem{Astrov:1960}
Astrov~D N.
\newblock The magnetoelectric effect in antiferromagnetics.
\newblock {\em Sov. Phys. JETP}, 11:708--709, 1960.

\bibitem{Astrov1961}
Astrov~D N.
\newblock Magnetoelectric effect in chromium oxide.
\newblock {\em Sov. Phys. JETP}, 13:729, 1961.

\bibitem{Brockhouse1953}
Brockhouse~B N.
\newblock Antiferromagnetic structure in {C}r$_2${O}$_3$.
\newblock {\em J. Chem. Phys.}, 21:961--962, 1953.

\bibitem{Dudko1971}
Dudko~K L, Eremenko~V V, and Semenenko~L M.
\newblock Magnetostriction of antiferromagnetic cr2o3 in strong magnetic
  fields.
\newblock {\em Phys. Status Solidi B}, 43:471--477, 1971.

\bibitem{Tobia2010}
Tobia D, De~Biasi E, Granada M, Troiani~H E, Zampieri G, Winkler E, and
  Zysler~R D.
\newblock Evolution of the magnetic anisotropy with particle size in
  antiferromagnetic {C}r$_2${O}$_3$ nanoparticles.
\newblock {\em J. Appl. Phys}, 108:104303, 11 2010.

\bibitem{Fechner2018}
Fechner M, Sukhov A, Chotorlishvili L, Kenel C, Berakdar J, and Spaldin~N A.
\newblock Magnetophononics: Ultrafast spin control through the lattice.
\newblock {\em Phys. Rev. Mater.}, 2:064401, 6 2018.

\bibitem{Iniguez:2008}
Iniguez J.
\newblock {First-Principles Approach to Lattice-Mediated Magnetoelectric
  Effects}.
\newblock {\em Phys. Rev. Lett.}, 101(11):117201, 2008.

\bibitem{blMalashevich2012}
Malashevich A, Coh S, I~Souza, and Vanderbilt D.
\newblock Full magnetoelectric response of {C}r$_2${O}$_3$ from first
  principles.
\newblock {\em Phys. Rev. B Condens.}, 86, 9 2012.

\bibitem{Mostovoy_et_al:2010}
Mostovoy M, Scaramucci A, Spaldin~N A, and Delaney~K T.
\newblock Temperature-dependent magnetoelectric effect from first principles.
\newblock {\em Phys. Rev. Lett.}, 105:087202, 2010.

\bibitem{Brown/Forsyth/Tasset:1998}
Brown~P J, Forsyth~J B, and Tasset F.
\newblock A study of magnetoelectric domain formation in {Cr$_2$O$_3$}.
\newblock {\em J. Phys.: Condens. Matter}, 10:663--672, 1998.

\bibitem{Brown/Forsyth/Tasset:1999}
Brown~P J, Forsyth~J B, and Tasset F.
\newblock Precision determination of antiferromagnetic form factors.
\newblock {\em Phys. B: Condens.}, 237:215--220, 1999.

\bibitem{Brown_et_al:2002}
Brown~P J, Forsyth~J B, Leli\`evre-Berna E, and Tasset F.
\newblock Determination of the magnetization distribution in {Cr$_2$O$_3$}
  using spherical neutron polarimetry.
\newblock {\em J. Phys.: Condens. Matter}, 14:1957–1966, 2002.

\bibitem{Forsyth_1988}
Tasset F, Brown~P J, and Forsyth~J B.
\newblock Determination of the absolute magnetic moment direction in cr2o3
  using generalized polarization analysis.
\newblock {\em J. Appl. Phys.}, 63(8):3606--3608, 1988.

\bibitem{Ye/Vanderbilt:2014}
Ye~M and Vanderbilt D.
\newblock Dynamical magnetic charges and linear magnetoelectricity.
\newblock {\em Phys. Rev. B}, 89(6):064301, 2014.

\bibitem{Bousquet/Spaldin/Delaney:2011}
Bousquet E, Spaldin~N A, and Delaney~K T.
\newblock Unexpectedly large electronic contribution to linear
  magnetoelectricity.
\newblock {\em Phys.\ Rev.\ Lett.}, 106:107202, 2011.

\bibitem{Gonze:2020}
Gonze X, Amadon B, Antonius G, Arnardi F, Baguet L, Beuken J-M, Bieder J,
  Bottin F, Bouchet J, Bousquet E, Brouwer N, Bruneval F, Brunin G, Cavignac T,
  Charraud J-B, Chen W, Côté M, Cottenier S, Denier J, Geneste G, Ghosez P,
  Giantomassi M, Gillet Y, Gingras O, Hamann~D R, Hautier G, He~X, Helbig N,
  Holzwarth N, Jia Y, Jollet F, Lafargue-Dit-Hauret W, Lejaeghere K, Marques
  M~A L, Martin A, Martins C, Miranda H~P C, Naccarato F, Persson K, Petretto
  G, Planes V, Pouillon Y, Prokhorenko S, Ricci F, Rignanese G-M, Romero~A H,
  Schmitt~M M, Torrent M, van Setten M~J, Van~Troeye B, Verstraete~M J, Zérah
  G, and J~W Zwanziger.
\newblock The {A}binit project: {I}mpact, environment and recent developments.
\newblock {\em Comput. Phys. Commun.}, 248:107042, 2020.

\bibitem{Romero:2020}
Romero~A H, Allan~D C, Amadon B, Antonius G, Applencourt T, Baguet L, Bieder J,
  Bottin F, Bouchet J, Bousquet E, Bruneval F, Brunin G, Caliste D,
  C{\^o}t{\'e} M, Denier J, Dreyer C, Ghosez P, Giantomassi M, Gillet Y,
  Gingras O, Hamann~D R, Hautier G, Jollet F, Jomard G, Martin A, Miranda H~P
  C, Naccarato F, Petretto G, Pike~N A, Planes V, Prokhorenko S, Rangel T,
  Ricci F, G-M Rignanese, Royo M, Stengel M, Torrent M, van Setten M~J,
  Van~Troeye B, Verstraete~M J, Wiktor J, Zwanziger~J W, and Gonze X.
\newblock {ABINIT}: Overview, and focus on selected capabilities.
\newblock {\em J. Chem. Phys.}, 152:124102, 2020.

\bibitem{ELK}
The {E}lk code : {A}n all-electron full-potential linearised augmented-plane
  wave ({LAPW}) code.
\newblock \url{https://elk.sourceforge.io/}, 2020.

\bibitem{Gianozzi_et_al:2009}
Giannozzi P, Baroni S, Bonini N, Calandra M, Car R, Cavazzoni C, Ceresoli D,
  Chiarotti~G L, Cococcioni M, Dabo I, Dal~Corso A, de~Gironcoli~S, Fabris S,
  Fratesi G, Gebauer R, Gerstmann U, Gougoussis C, Kokalj A, Lazzeri M,
  Martin-Samos L, Marzari N, Mauri F, Mazzarello R, Paolini S, Pasquarello A,
  Paulatto L, Sbraccia C, Scandolo S, Sclauzero G, Seitsonen~A P, Smogunov A,
  Umari P, and Wentzcovitch~R M.
\newblock {QUANTUM} {ESPRESSO}: a modular and open-source software project for
  quantum simulations of materials.
\newblock {\em J. Phys. Condens. Matter}, 21:395502, 2009.

\bibitem{Giannozzi_et_al:2017}
Giannozzi P, Andreussi O, Brumme T, Bunau O, Buongiorno~Nardelli M, Calandra M,
  Car R, Cavazzoni C, Ceresoli D, Cococcioni M, Colonna N, Carnimeo I,
  Dal~Corso A, de~Gironcoli~S, Delugas P, DiStasio~R A, Ferretti A, Floris A,
  Fratesi G, Fugallo G, Gebauer R, Gerstmann U, Giustino F, Gorni T, Jia J,
  Kawamura M, Ko~H-Y, Kokalj A, Küçükbenli E, Lazzeri M, Marsili M, Marzari
  N, Mauri F, Nguyen~N L, Nguyen H-V, Otero de-la Roza~A, Paulatto L, Poncé S,
  Rocca D, Sabatini R, Santra B, Schlipf M, Seitsonen~A P, Smogunov A, Timrov
  I, Thonhauser T, Umari P, Vast N, Wu~X, and Baroni S.
\newblock Advanced capabilities for materials modelling with {Q}uantum
  {ESPRESSO}.
\newblock {\em J. Phys.:Condens. Matter}, 29(46):465901, oct 2017.

\bibitem{Kresse/Furthmueller_CMS:1996}
Kresse G and Furthm{\"u}ller J.
\newblock Efficiency of ab-initio total energy calculations for metals and
  semiconductors using a plane-wave basis set.
\newblock {\em Comput. Mater. Sci.}, 6:15--50, 1996.

\bibitem{Kresse/Furthmueller_PRB:1996}
Kresse G and Furthm{\"u}ller J.
\newblock Efficient iterative schemes for ab initio total-energy calculations
  using a plane-wave basis set.
\newblock {\em Phys.\ Rev.\ B}, 54:11169--11186, 1996.

\bibitem{Bloechl:1994}
Bl{\"o}chl~P E.
\newblock Projector augmented-wave method.
\newblock {\em Phys.\ Rev.\ B}, 50:17953--17979, 1994.

\bibitem{PriComm_delayney_2023}
Private communication, 2023.

\bibitem{LelievreBerna:2007tm}
Leli{\`e}vre-Berna E, Brown~P J, Tasset F, Kakurai K, Takeda M, and Regnault
  LP.
\newblock {Precision manipulation of the neutron polarisation vector}.
\newblock {\em Phys. B: Condens. Matter}, 397(1-2):120--124, July 2007.

\bibitem{Tasset:1999wp}
Tasset F, Brown~P J, E~Leli{\`e}vre-Berna, Roberts~T W, Pujol S., Allibon J,
  and Bourgeat-Lami E.
\newblock {Spherical neutron polarimetry with Cryopad-II}.
\newblock {\em Phys. B: Condens.}, 267-268:69--74, 1999.

\bibitem{LelievreBerna:2005vj}
Leli{\`e}vre-Berna E, Bourgeat-Lami E, Fouilloux P, Geffray B, Gibert Y,
  Kakurai K, Kernavanois N, Longuet B, Mantegazza F, Nakamura M, Pujol S,
  Regnault LP, Tasset F, Takeda M, Thomas M, and Tonon X.
\newblock {Advances in spherical neutron polarimetry with Cryopad}.
\newblock {\em Phys. B: Condens. Matter}, 356(1-4):131--135, February 2005.

\bibitem{Mag2pol}
Qureshi N.
\newblock {{\it Mag2Pol}: a program for the analysis of spherical neutron
  polarimetry, flipping ratio and integrated intensity data}.
\newblock {\em J. Appl. Crystallogr.}, 52(1):175--185, Feb 2019.

\bibitem{He_et_al:2010}
He~X, Wang Y, Wu~N, Caruso~A N, Vescovo E, Belashchenko~K D, Dowben~P A, and
  Binek C.
\newblock {Robust isothermal electric control of exchange bias at room
  temperature}.
\newblock {\em Nat. Mater.}, 9:579--585, 2010.

\bibitem{Borisov2016}
Borisov P, Ashida T, Nozaki T, Sahashi M, and Lederman D.
\newblock Magnetoelectric properties of 500-nm cr2o3 films.
\newblock {\em Phys. Rev. B}, 93:174415, 5 2016.

\bibitem{Weber2023}
Weber~S F and Spaldin~N A.
\newblock Characterizing and overcoming surface paramagnetism in
  magnetoelectric antiferromagnets.
\newblock {\em Phys. Rev. Lett.}, 130:146701, 4 2023.

\bibitem{Perdew/Zunger:1981}
Perdew~J A and Zunger A.
\newblock Self-interaction correction to density-functional approximations for
  many-electron systems.
\newblock {\em Phys. Rev. B}, 23:5048--5079, 1981.

\bibitem{Liechtenstein_U}
Liechtenstein~A I, Anisimov~V I, and Zaanen J.
\newblock {Density-functional theory and strong interactions: Orbital ordering
  in Mott-Hubbard insulators}.
\newblock {\em Phys. Rev. B}, 52(8):R5467, 1995.

\bibitem{shi:2009}
Shi S, Wysocki~A L, and Belashchenko~K D.
\newblock Magnetism of chromia from first-principles calculations.
\newblock {\em Phys. Rev. B}, 79(10):104404, March 2009.

\bibitem{mu:2014}
Mu~S, Wysocki~A L, and Belashchenko~K D.
\newblock First-principles microscopic model of exchange-driven magnetoelectric
  response with application to {C}r$_2${O}$_3$.
\newblock {\em Phys. Rev. B}, 89(17):174413, May 2014.

\bibitem{Monkhorst/Pack:1976}
Monkhorst~H J and Pack~J D.
\newblock Special points for brillouin-zone integrations.
\newblock {\em Phys.\ Rev.\ B}, 13:5188--5192, 1976.

\bibitem{phonopy}
Togo A and Tanaka I.
\newblock First principles phonon calculations in materials science.
\newblock {\em Scr. Mater.}, 108:1--5, Nov 2015.

\bibitem{phonopy-phono3py-JPSJ}
Togo A.
\newblock First-principles phonon calculations with phonopy and phono3py.
\newblock {\em J. Phys. Soc. Jpn.}, 92(1):012001, 2023.

\bibitem{Hill:2010}
Hill~A H, Harrison A, Dickinson C, W~Zhou, and Kockelmann W.
\newblock Crystallographic and magnetic studies of mesoporous eskolaite,
  {C}r$_2${O}$_3$.
\newblock {\em Microporous and Mesoporous Materials}, 130(1):280--286, 2010.

\bibitem{thermo_pw}
\texttt{thermo\_pw} is an extension of the \uppercase{Q}uantum
  \texttt{ESPRESSO} (\uppercase{QE}) package which provides an alternative
  organization of the \uppercase{QE} workflow for the most common tasks.
  \uppercase{F}or more information see
  \url{https://dalcorso.github.io/thermo_pw/}.

\bibitem{Perdew/Burke/Ernzerhof:1996}
Perdew~J P, Burke K, and Ernzerhof M.
\newblock Generalized gradient approximation made simple.
\newblock {\em Phys. Rev. Lett.}, 77(18):3865--3868, Oct 1996.

\bibitem{US_Vanderbilt}
Vanderbilt D.
\newblock {Soft self-consistent pseudopotentials in a generalized eigenvalue
  formalism}.
\newblock {\em Phys. Rev. B}, 41(11):7892, 1990.

\bibitem{pslibrary}
Dal~Corso A.
\newblock {Pseudopotentials periodic table: From H to Pu}.
\newblock {\em Comp. Mater. Sci.}, 95:337, 2014.

\bibitem{pslibrary_2}
See \url{https://dalcorso.github.io/pslibrary/}.

\bibitem{DFPT_review_Baroni}
Baroni S, de~Gironcoli~S, Dal~Corso A, and Giannozzi P.
\newblock Phonons and related crystal properties from density-functional
  perturbation theory.
\newblock {\em Rev. Mod. Phys.}, 73:515--562, Jul 2001.

\bibitem{pseudodojo}
Van~Setten MJ, Giantomassi M, Bousquet E, Verstraete~M J, Hamann~D R, Gonze X,
  and Rignanese G-M.
\newblock The pseudodojo: Training and grading a 85 element optimized
  norm-conserving pseudopotential table.
\newblock {\em Comput. Phys. Commun.}, 226:39, 2018.

\end{thebibliography}

\end{document}